\documentclass{revtex4-1}
\usepackage{bm}
\usepackage{graphicx}
\usepackage{amsfonts,amssymb,amsmath}
\usepackage{comment}
\usepackage{color}
\usepackage{esint}

\newcommand{\rmT}[0]{\textrm{T}}
\newcommand{\rmi}[0]{\textrm{i}}
\newcommand{\rme}[0]{\textrm{e}}

\newcommand{\red}[1]{\textcolor{black}{#1}}
\newcommand{\phasediff}[0]{\Delta \phi}

\begin{document}
\title{Direct extraction of phase dynamics from fluctuating rhythmic data based on a Bayesian approach}
\date{\today}

\author{Kaiichiro Ota}
\email[]{kaiichiro@acs.i.kyoto-u.ac.jp}
\affiliation{Graduate School of Informatics, Kyoto University, Kyoto 606-8501, Japan}
\affiliation{JST CREST, Sanbancho, Chiyoda-ku, Tokyo 102-0075, Japan}

\author{Toshio Aoyagi}
\email[]{aoyagi@i.kyoto-u.ac.jp}
\affiliation{Graduate School of Informatics, Kyoto University, Kyoto 606-8501, Japan}
\affiliation{JST CREST, Sanbancho, Chiyoda-ku, Tokyo 102-0075, Japan}

\begin{abstract}
	Employing both Bayesian statistics and the theory of
	nonlinear dynamics, we present a practically efficient method to
	extract a phase description of weakly coupled limit-cycle
	oscillators directly from time series observed in a rhythmic system.
	As a practical application, we numerically demonstrate that this method can
	retrieve all the interaction functions from the
	fluctuating rhythmic neuronal activity exhibited by a network of asymmetrically
	coupled neurons.  This method can be regarded as a type of statistical phase
	reduction method that requires no detailed modeling, 
	and as such, it is a very practical and reliable method in application
	to data-driven studies of rhythmic systems.
\end{abstract}

\maketitle

Theoretical models have provided great insight
into the nature of real-world dynamic phenomena \cite{Hoppensteadt:1997tp,Alon:2006tm,Nowak:2006wy,Barrat:2008wp}.
In general, to understand some phenomena of interest, 
we need to construct a good theoretical 
model that accounts for experimental data. 
Successful theoretical models can be roughly divided
into two classes. One class consists of detailed models constructed to faithfully reproduce
as many characteristics of the systems under study as possible.
Such models contribute to the quantitative understanding of the dynamical behavior
of the specific systems to which they are applied. 
The other class consists of abstract models constructed to capture
some essential aspect of the systems of interest, such as rhythmic behavior.  This
type of model is not intended to accurately simulate all the dynamical behavior
of a specific system, but rather to provide a description of some universal aspect of its
dynamics. 
The advantage of this type of model is that, because it does not focus on 
the detailed behavior of any specific system, but rather on the universal aspects
of this behavior, it can provide a unified framework for describing the behavior
exhibited by a wide range of dynamical systems.
In this way, such models allow us to gain a deeper understanding of the universal
mechanisms existing in broad classes of systems.

\begin{figure}[b]
	\includegraphics[width=8cm]{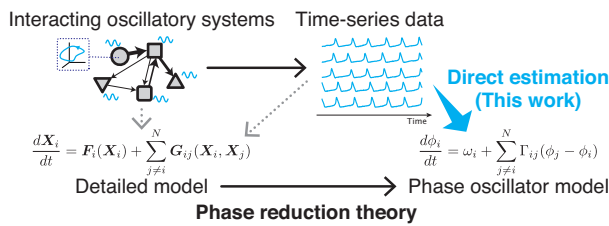}
	\caption{
		Conceptual diagram of the method for obtaining phase dynamics from
		observed rhythmic time series. In the conventional approach, we first
		construct a detailed model based on experimental observations and next
		obtain the phase oscillator model by applying the phase reduction method
		to this detailed model. Our proposed alternative approach begins with a 
		phase oscillator model of generic form.
		We then determine through use of a Bayesian statistical method the explicit
		content of this model directly from the observed time-series data,
		without constructing the detailed model. This can be regarded as a statistical 
		phase reduction method based on Bayesian theory.
	}
	\label{fig:concept}
\end{figure}

One successful model of the abstract type described above is the phase description of
the dynamics of interacting oscillatory systems
(Fig.~\ref{fig:concept}).  In such a model, the evolution of each oscillatory
system is described by a single degree of freedom, the
phase.  In this description, the dynamics of a system of $N$ coupled oscillators
is generally described by a set of equations of the form
\begin{equation}
	\frac{d\phi_i}{dt} = \omega_i+ \sum^N_{j \neq i}\Gamma_{ij}(\phi_j-\phi_i) \quad 
	\red{(i=1,\dots,N)},
	\label{phaseeq}
\end{equation}
where $\phi_i$ is the phase of the $i$-th oscillator, representing the
timing of its oscillation \cite{Kuramoto:1984wo}.  The parameter $\omega_i$ and
the function $\Gamma_{ij}(\phasediff)$ denote the natural frequency of the
$i$-th oscillator and the coupling function from the $j$-th oscillator to $i$-th
oscillator, respectively.
These coupling functions and natural frequencies can be theoretically determined 
using a detailed model of the form
\red{$\frac{d\bm{X}_i}{dt} = \bm{F}_i(\bm{X}_i) + \sum^N_{j \neq i}\bm{G}_{ij}(\bm{X}_i,\bm{X}_j)$} (where \red{$\bm{X}_i$} denotes the multidimensional 
state of the \red{$i$-th oscillator}),
whose dynamics generally have a large number of degrees of freedom.
In fact, it has been found that, employed in this manner, 
the theory of dynamical systems allows for the construction of models of the form
(\ref{phaseeq}) that provide descriptions of a broad class of systems of limit-cycle 
oscillators.
Specifically, this class consists of those systems in which the interactions between 
oscillators only affect the phase asymptotically.
In particular, it has been found that rhythmic systems of diverse types can be treated
by models of this form \cite{Strogatz:2003we}.

The conventional method for constructing the 
phase oscillator model for a specific system consists of two steps:
In the first, a detailed model is constructed from experimental data,
and in the second,
a phase model of the form (\ref{phaseeq}) is derived from this detailed model 
by applying the phase reduction theory (Fig.~\ref{fig:concept}).
However, it is often unfeasible to construct the correct detailed model in one step,
because the underlying dynamics are generally nonlinear and of high dimension.
For this reason, derivation of the phase model using the conventional approach is often
quite complicated and time consuming.
In this Letter, we propose an alternative approach to
describing the dynamics of such a network that forgoes
the detailed model used in the conventional approach and,
instead, begins with (\ref{phaseeq}), slightly generalized to include a noise term as
\begin{equation}
	\frac{d\phi_i}{dt} = \omega_i + \sum_{j \neq i}^N \Gamma_{ij}(\phi_j-\phi_i) + \eta_i(t).
	\label{phasenoise}
\end{equation}
In our approach, we skip the first step of the conventional approach and
determine the explicit content of the phase model given in (\ref{phasenoise}) 
directly from time-series data (Fig.~\ref{fig:concept}).
This can be regarded as a statistical version of the phase reduction method based on Bayesian theory.
Here, we introduce the noise $\eta_i(t)$, which represents an
unavoidable source of uncertainty, for example,
arising from observational error.  
For simplicity, we assume that each noise function $\eta_i(t)$ is
independent Gaussian white noise satisfying $\langle \eta_i(t) \rangle = 0,
\langle \eta_i(t)\eta_j(s) \rangle = 2D_i \delta_{ij} \delta(t-s)$,
where $\delta_{ij}$ is the Kronecker delta and $D_i$ represents the
strength of the noise.  Thus, in the approach we propose, we must determine the nonlinear
coupling functions $\Gamma_{ij}(\phasediff)$ and the model parameters
$\omega_i$ and $D_i$ in the dynamical system (\ref{phasenoise}) so as to best predict the
dynamical behavior of the observed system.  
This is a typical nonlinear optimization problem.
Such problems are often difficult to treat because there generally exist
many local optimal solutions, owing to the nonlinearity.
To overcome this difficulty, we employ a Bayesian statistical approach, which
allows us to derive the phase oscillator model directly from the time-series data
\cite{Shandilya:2011dh,Cadieu:2010vn,Tokuda:2007uj,Kralemann:2007wp,Kralemann:2008gh,Kralemann:2011fj,Stankovski:2012el,Duggento:2012el}.

Consider the situation in which we observe $N$ oscillatory signals, $s_i(t)\,(i=1,\dots,N)$,
each of which is generated by a separate limit-cycle oscillator,
and suppose that these oscillators are weakly coupled.
Further, we assume that each signal $s_i(t)$ is sampled at $T+1$ discrete time points
\red{$t_\tau = t_1 + (\tau-1) \Delta t$, where $\tau=1,2,\dots,T+1$} and $\Delta t$ is the sampling interval.

\begin{figure}[b]
	\includegraphics{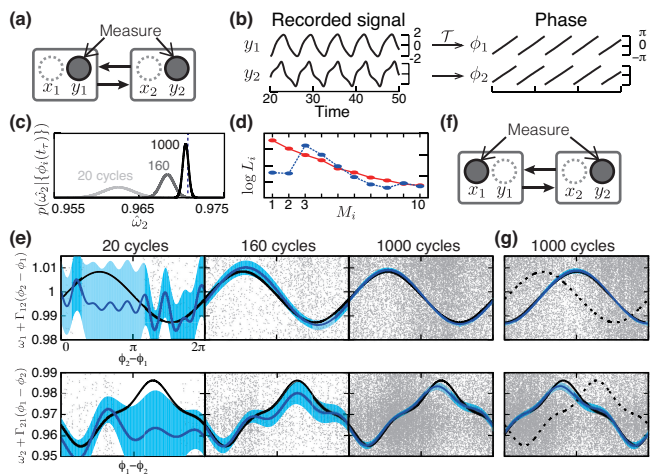}
	\caption{ 
		Phase description extracted directly from rhythmic signals
		in a system of two mutually coupled van der Pol oscillators \red{\cite{SM}.}
		\red{From the two state variables of each oscillator,}
		only the variable $y_i(t)$ is observed as a signal, as depicted in (a).  
		(b) The recorded signals $y_i(t)$ (left graphs) are transformed
		into phase time series $\phi_i(t)$ (right graphs). Both $y_i(t)$ and $\phi_i(t)$
		fluctuate slightly, owing to the interactions and noise, though the fluctuations
		are difficult to discern here.
		Typical data taken over approximately five oscillation cycles are plotted.
		(c) Posterior probability density distribution of $\hat{\omega}_2$ 
		calculated for three observation durations\red{.} 
		\red{The dashed vertical line represents the theoretical value.}
		(d) Log marginal likelihood for various values of $M_i$ in the case of the data measured with 
		1000 oscillation cycles.
		Following the Bayesian model selection, we choose the value $M_i$ giving the largest
		marginal likelihood. The log marginal likelihoods for $\hat{\Gamma}_{12}$ (red solid curve) and
		$\hat{\Gamma}_{21}$ (blue dashed curve) have maximum values at $M_1^* =1$ and $M_2^* =3$,
		respectively. 
		(e) Posterior distributions for the deterministic terms 
		$\omega_i+\Gamma_{ij}(\Delta \phi)$.
		The blue curves and the light blue regions represent
		the mean and 95\% confidence interval of the distribution, 
		respectively.
		The black curves are those obtained from the phase reduction theory,
		and the numerous gray dots denote data points.
		The graphs in each column display the results obtained using the data measured with the three 
		different time durations\red{.} 
		(f,g) Here, it is seen that even if different types of dynamical variables 
		(i.e., \red{$x_1$ and $y_2$})
		are used as the signals (panel f), 
		the phase coupling functions can still be reliably retrieved,
		except for an inevitable uncertainty in the phase shift (see panel g and the main text).
		For comparison, the dashed curves represent the theoretical curves with the correct phase relationship.
	}
	\label{fig:vdp-N2}
\end{figure}

Our method consists of two main steps．
In the first step, we transform each observed signal into a time series
of the phase (Fig.~\ref{fig:vdp-N2}b). For this purpose, using the Hilbert
transformation $s_i^\mathcal{H}(t)$ of the signal $s_i(t)$, we
construct a prototype of the phase $\theta_i(t)$, as defined by
$A_i(t)\rme^{\rmi \theta_i(t)} = s_i(t) + \rmi s_i^\mathcal{H}(t)$ \cite{Pikovsky:2004vm}. 
However, the variable $\theta_i$ here generally
differs from the phase $\phi_i$ used in the phase description
(\ref{phasenoise}), because $\theta_i$ does not increase
with time at a constant rate in the absence of both interactions and noise.  In the
context of dynamical systems described by (\ref{phasenoise}), the
phase should be chosen as a quantity that changes in time at a constant rate 
in the absence of noise and interactions.
Using the fact that the probability density distribution of $\theta_i$,  $f(\theta_i)$, is inversely
proportional to $d\theta_i/dt$ statistically, Kralemann {\it et al.}~proposed the following
transformation from the prototype phase $\theta_i$ to the phase $\phi_i$:
$\phi(\theta) = 2\pi \int_0^\theta f(\theta') d\theta'$
\cite{Kralemann:2007wp,Kralemann:2008gh,Kralemann:2011fj}.  
With the above two procedures, we can transform the observed signals $s_i(t_\tau)$
into $N$ time series of the phase $\phi_i(t_\tau)$ $(\tau=1,2,\cdots,T+1)$, which are
expected to increase linearly with time in the absence of noise and interactions.
In general, the presence of noise and interactions causes slight fluctuations of the phases.
These fluctuations contain information from which the explicit content of (\ref{phasenoise})
can be inferred.

As an illustrative example, we applied the method described above to 
a system of two coupled van der Pol oscillators, as shown in Fig.~\ref{fig:vdp-N2}.
For the parameter values used there, each oscillator exhibits limit-cycle oscillation 
in the absence of coupling. Here we assume that for the $i$-th oscillator only the time-series of the
variable $y_i$ can be observed as the signal, $s_i(t)$ (i.e., $x_i$ is unobservable),
as illustrated in Fig.~\ref{fig:vdp-N2}a.
Figure \ref{fig:vdp-N2}b exhibits a typical result of the transformation from the 
signals $y_i$ to the phases $\phi_i$\red{.}

The second step is to determine the explicit content of the phase oscillator model (\ref{phasenoise}) needed to
generate the obtained phase time-series $\phi_i(t_\tau)$. First, we specify all of the 
parameters to be evaluated.
The $2\pi$-periodic coupling function $\Gamma_{ij}$ can be expanded
in a Fourier series as
$\Gamma_{ij}(\phasediff) = a_{ij}^{(0)} + \sum_{m=1}^{M_i} \left[ a_{ij}^{(m)} \cos(m\phasediff) + b_{ij}^{(m)} \sin(m\phasediff) \right]$.
In this expansion, we keep only up to the $M_i$-th harmonic for each $\Gamma_{ij}$.
The parameters $M_i$ control the complexity of the model and 
can be determined using a model selection method, as explained below.
Except in certain particular situations, $\{a^{(0)}_{ij}\}_{j \neq i}$ and $\omega_i$ are redundant,
because their contributions to the dynamics are inseparable \cite{Kralemann:2008gh}.
We therefore treat $\hat{\omega}_i \equiv \omega_i + \sum_{j \neq i}^N a_{ij}^{(0)}$
as a single parameter.
Thus, in all, we must evaluate $2+2M_i(N-1)$ unknown model parameters,
$\hat{\omega}_i$, $D_i$ and $\{a_{ij}^{(m)}, b_{ij}^{(m)}\}_{m,j}$. 
For simplicity, hereafter we use the shorthand notation 
$\bm{c}_i \equiv [\hat{\omega}_i, \bm{c}_{i,1}, \cdots, \bm{c}_{i,i-1},\bm{c}_{i,i+1},\cdots, \bm{c}_{i,N}]^\rmT$,
with 
$\bm{c}_{i,j} \equiv \left[a_{ij}^{(1)}, b_{ij}^{(1)}, a_{ij}^{(2)}, b_{ij}^{(2)}, \cdots, a_{ij}^{(M_i)}, b_{ij}^{(M_i)}\right]$ and $\psi_{ij} \equiv \phi_j - \phi_i$.

We next evaluate the above parameters from the phase time-series 
$\{\phi_i(t_\tau)\}$ $(i=1,\dots,N; \tau=1,\dots,T+1)$
on the basis of the Bayesian statistical framework \cite{bishop2006,Murphy:2012uq}.
First, we write the probability to reproduce the observed phase time series $\{\phi_i(t_\tau)\}$
given $\bm{c}_i$ and $D_i$ as
\begin{equation}
	p(\{\phi_i(t_\tau)\}|\bm{c}_i,D_i) 
	= \prod_{\tau=1}^T \mathcal{N}\left( \hat{\omega}_i 
	+ \sum_{j \neq i}^N \hat{\Gamma}_{ij}[\psi_{ij}(t_\tau)], \sigma_i^2 \right),
	\label{likelihood}
\end{equation}
where $\hat{\Gamma}_{ij}(\phasediff) \equiv \Gamma_{ij}(\phasediff)-a_{ij}^{(0)}$
and $\sigma^2_i \equiv \frac{2D_i}{\Delta t}$.
Here, $\mathcal{N}(\mu,\sigma^2)$ denotes the density of the Gaussian distribution 
with mean $\mu$ and variance $\sigma^2$.
Next, following the standard Bayesian approach, we introduce a probability density 
distribution of the unknown parameters written $p(\bm{c}_i,D_i)$, which allows us to 
compute not only the most probable parameter values (maximum likelihood estimates)
but also their uncertainties.
When we obtain new observed data $\{\phi_i(t_\tau)\}$, 
the parameter distribution $p(\bm{c}_i,D_i)$ is updated according to Bayes' theorem,
\begin{equation}
	p(\bm{c}_i,D_i|\{\phi_i(t_j)\})\propto p(\{\phi_i(t_j)\}|\bm{c}_i,D_i)p(\bm{c}_i,D_i),
	\label{Bayes}
\end{equation}
where $p(\bm{c}_i,D_i)$ and $p(\bm{c}_i,D_i|\{\phi_i(t_j)\})$ are called the ``prior''
and ``posterior'' distributions, respectively.
Although the choice of the functional form of the prior distribution is somewhat 
arbitrary, it is convenient to use a conjugate prior distribution so that 
the posterior distribution derived from (\ref{Bayes}) has the same functional 
form as the prior distribution. In particular, if the conjugate prior distribution can be characterized by
some parameters (called hyperparameters), we have only to update the values of the 
hyperparameters to obtain the posterior distribution. 
For the conjugate prior distribution, we adopt a Gaussian-inverse-gamma distribution \cite{SM}, given by
\begin{equation}
	p(\bm{c}_i,D_i) \propto 
	e^{ -\frac{(\bm{c}_i-\bm{\chi}_i)^\rmT \Sigma_i^{-1} (\bm{c}_i-\bm{\chi}_i) + 2\beta_i}{2\sigma_i^2} } 
	(\sigma_i^2)^{-\frac{P_i}{2} -\alpha_i -1},
	\label{prior}
\end{equation}
where $P_i$ is the dimension of the vector $\bm{c}_i$. 
Note that the prior distribution for $\bm{c}_i$ and $D_i$ is characterized fully by 
the hyperparameters $\bm{\chi}_i, \Sigma_i, \alpha_i$ and $\beta_i$.
Using Eq.~(\ref{Bayes}) with Eqs.~(\ref{likelihood}) and (\ref{prior}),
we can easily compute the hyperparameters of the posterior distribution as 
follows:
\begin{eqnarray*}
	\bm{\chi}^\mathrm{new}_i &=& \Sigma^\mathrm{new}_i (F_i^\rmT \bm{\delta}_i + (\Sigma_i^\mathrm{old})^{-1} \bm{\chi}_i^\mathrm{old}),\\
	\Sigma^\mathrm{new}_i &=& \left\{(\Sigma_i^\mathrm{old})^{-1} + F_i^\rmT F_i \right\}^{-1},\\
	\alpha^\mathrm{new}_i &=& \alpha_i^\mathrm{old} + \frac{T}{2},\\
	\beta^\mathrm{new}_i  &=& \beta^\mathrm{old}_i + \frac{1}{2} \{ \bm{\delta}_i^\rmT \bm{\delta}_i + (\bm{\chi}_i^\mathrm{old})^\rmT (\Sigma_i^\mathrm{old})^{-1} \bm{\chi}^\mathrm{old}_i\\
	&\quad& \quad \quad - (\bm{\chi}_i^\mathrm{new})^\rmT (\Sigma_i^\mathrm{new})^{-1} \bm{\chi}^\mathrm{new}_i\}.
	\label{update}
\end{eqnarray*}
Here we have defined the $T$-dimensional column vectors
$(\bm{\delta}_i)_\tau \equiv \frac{\phi_i(t_{\tau+1})-\phi_i(t_\tau)}{\Delta t} \ (\tau=1,\dots,T)$
and the $T \times P_i$ matrices
\begin{equation*}
	\red{
	F_i = \begin{bmatrix}
		1 & \bm{G}^{1}_{i,1} & \cdots & \bm{G}^{1}_{i,i-1} & \bm{G}^{1}_{i,i+1} & \cdots & \bm{G}^{1}_{i,N} \\
		&              &        &    \vdots      &                &        & \\
		1 & \bm{G}^{T}_{i,1} & \cdots & \bm{G}^{T}_{i,i-1} & \bm{G}^{T}_{i,i+1} & \cdots & \bm{G}^{T}_{i,N} 
	\end{bmatrix},
}
\end{equation*}
with the $M_i$-dimensional row vectors
$\bm{G}^{\tau}_{i,j} \equiv [
	\cos \psi_{ij}(t_\tau), \sin \psi_{ij}(t_\tau), 
	\cos (2\psi_{ij}(t_\tau)), \sin (2\psi_{ij}(t_\tau)), \cdots, \\
\cos (M_i\psi_{ij}(t_\tau)), \sin (M_i\psi_{ij}(t_\tau)) ].$
The superscripts ``new'' and ``old'' indicate the hyperparameters of the posterior and 
prior distributions, respectively.

In the case of van der Pol oscillators, a typical form of the posterior distribution for 
$\hat\omega_2$ is displayed in Fig.~\ref{fig:vdp-N2}c.  The different curves correspond to
posterior distributions obtained using observations with different durations.  
It is seen that as the amount of data is increased,
the peak of the posterior distribution becomes sharper and
closer to the theoretical value (dashed vertical line).
This implies that
the estimated mean value becomes both more accurate and more precise as the
amount of data increases.

Now we return to the determination of $M_i$, which controls
the degree of approximation of $\hat{\Gamma}_{ij}(\phasediff)$
resulting from the truncation of the Fourier series.  
Bayesian theory provides an effective method to choose the ``best'' model with certain values of
the parameters $M^*_i$.  In this method, using the posterior distributions calculated
with Eq.~(\ref{likelihood}) for various values of $M_i=1,2,\cdots$, we
evaluate the values of the marginal likelihood functions
$L_i(M_i)\equiv\dotsint p(\bm{\phi}_i|\bm{c}_i, D_i, M_i) d\bm{c}_i \,
dD_i$.  It is well known that the quantities $L_i$ measure the goodness
of a fit over all possible values of the parameters, taking account
of the model's complexity, which is essentially given by the total
number of model parameters to be evaluated \cite{bishop2006,Murphy:2012uq}.
Then, it is reasonable to choose the optimal value $M_i^*$ for each $M_i$ such that
$L_i(M_i^*) = \max_{M_i}\{L_i(M_i)\}$ \cite{SM}.

The dependence of the marginal likelihood function on $M_i$ for the
case of van der Pol oscillators is plotted in Fig.~\ref{fig:vdp-N2}d.
Note that $L_i$ generally tends to decrease as a function of $M_i$ for sufficiently large
$M_i$, because the number of free parameters is too large (i.e., the model is too complex).
The graph shows that $L_1$ and $L_2$ are maximal at $M_1 = 1$ and $M_2
= 3$, respectively.  This result implies that the function
$\hat{\Gamma}_{12}(\phasediff)$ is accurately approximated by only the first
Fourier mode, whereas we need to consider up to the third mode for
$\hat{\Gamma}_{21}(\phasediff)$.  Using the posterior distribution
obtained with $M_1 = 1$ and $M_2 = 3$, we can calculate the posterior
density distribution for the functional form of the deterministic
terms $\omega_i + \Gamma_{ij}(\psi_{ij})$, as indicated in
Fig.~\ref{fig:vdp-N2}e.  It is seen that the estimated functions
converge to the theoretical ones as the amount of data increases.

It is somewhat surprising that, even if we use a pair of different state
variables $x_1$ and $y_2$ as the signal sources (Fig.~\ref{fig:vdp-N2}f), the
result of the estimation is essentially unchanged (Fig.~\ref{fig:vdp-N2}g). 
This suggests that the result is largely insensitive to the choice of the
observed signals;
in other words, we can use any variables that reflect the rhythmic
behavior of each oscillator.  This suggests that our method should be widely
applicable in various experimental settings.
We note that, as shown in Fig.~\ref{fig:vdp-N2}g,
an uncertainty in the phase shift is inevitable,
because even in principle we cannot know the phase relationship 
between $x_1$ and $y_2$ only from the data.  
However, it is seen that, other than such an inevitable phase shift,
the estimation is reasonably accurate.

\begin{figure}[b]
	\includegraphics[]{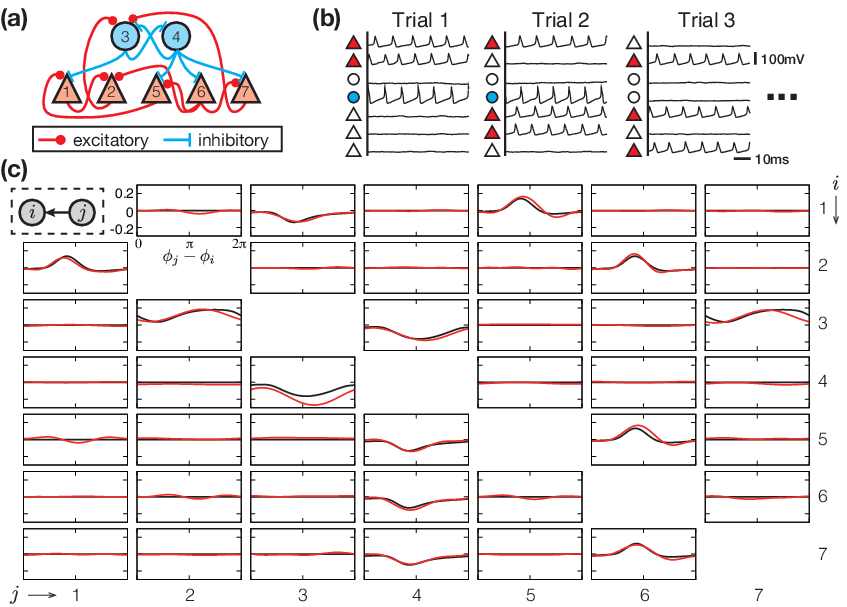}
	\caption{Extracting phase dynamics from the membrane voltages generated in 
		a network of spiking neurons.
		(a) Synaptic connections in a network consisting of 
		five excitatory and two inhibitory neurons \cite{*[{See Supplemental Material }] [{ for details concerning the method and additional results.}] SM}.
		(b) Experimental conditions for measuring the neuronal membrane voltages.
		In each trial, 3 or 4 randomly selected neurons fire, while the others are quiescent.
		In each case, the selected neurons are injected with different fixed levels of input current,
		and thus exhibit firing activity at different rates in the range 99--107 Hz. 
		The data set used here consists of 100 trials that contain 3700--4600 spikes for each cell.  
		The time duration of each trial is 1000 ms. 
		(c) Estimated mean curves (red) and theoretical curves (black) for all $\Gamma_{ij}$.
	}
	\label{fig:SN-N7}
\end{figure}

We next apply our method to a more practical case, in which a
network consisting of \red{a larger number of} synaptically coupled inhibitory and excitatory
neurons is investigated (Fig.~\ref{fig:SN-N7}a), and the neuronal
membrane voltages $V_i(t)$ are measured as signals.  In addition, we
consider more general and realistic experimental conditions,
in which only some of the neurons are actively spiking, while others
are inactive, as shown in Fig.~\ref{fig:SN-N7}b.  
In this treatment, we assume that
the properties of the synaptic connections do not change throughout
the measurement procedure\red{.}
In each trial, we randomly choose only three or four neurons to be activated by
injecting the selected neurons with different neuron-specific levels
of current.  One of the characteristics of the synaptic interaction
used here is that inactive neurons are not involved in the interaction
and thus have no effect on the dynamics of the system. This means that
in a single trial,
we can retrieve information only regarding the interactions
among the neurons that are active in that trial. 

Even with the information limited in the manner described above,
by combining the data from 
\red{sufficiently many experimental trials,}
we found that with our proposed method, we are able to obtain a phase description
directly from the observed time-series data.  Figure~\ref{fig:SN-N7}c
displays the result for the estimated mean of the coupling functions
$\Gamma_{ij}(\phasediff)$. 
We find that the estimated coupling functions are sufficiently close to the theoretical ones
that the resulting model correctly discriminates among inhibitory, excitatory and null couplings.
It is thus seen that the essential functional aspects of the neuronal network can be 
reconstructed from the voltage time-series data alone.
Furthermore, we have confirmed that the
coupling functions can also be successfully evaluated in the case of a
larger network \cite{SM}. 

In this Letter, we have proposed an approach for constructing a coupled phase oscillator
description of rhythmic behavior directly from fluctuating time-series data.
This approach combines the theory of 
nonlinear dynamics and a Bayesian statistical method.
We have demonstrated that this approach allows us to reconstruct the coupling functions
in a quantitatively accurate manner,
even in the case that only one of the states of each oscillator exhibiting rhythmic behavior is
measured.
Furthermore, we have confirmed that both the precision and accuracy of the 
reconstruction are essentially the same 
in the case that the observed state variables of the oscillators are of different types 
as in the case that they are of the same type.
We thus believe that our method will be quite useful
in application to actual experiments and that it will contribute to
data-driven studies of various rhythmic phenomena found in biological,
physical and social systems.

\begin{acknowledgments}
	We thank Y.~Iba for useful discussions.
	This work was supported by KAKENHI 25115719.
\end{acknowledgments}

\clearpage
\renewcommand{\thefigure}{S\arabic{figure}}
\setcounter{figure}{0}

\section*{Supplemental Material}
\section{The prior distribution and its hyperparameters}
The full form of a Gaussian-inverse-gamma distribution (Eq.~5 in the main text)
is written as
\begin{align*}
	p(\bm{c},D|\bm{\chi},\Sigma,\alpha,\beta) 
	&= \frac{1}{(2\pi\sigma^2)^{P/2}|\Sigma|^{1/2}} \exp \left\{ -\frac{1}{2\sigma^2}(\bm{c} - \bm{\chi})^\rmT \Sigma^{-1} (\bm{c} - \bm{\chi}) \right\} \frac{\beta^{\alpha}}{\Gamma(\alpha)} (\sigma^2)^{-\alpha-1} \exp \left( -\frac{\beta}{\sigma^2} \right),
\end{align*}
where $\sigma^2 \equiv \frac{2D}{\Delta t}$, and we omit the subscripts $i$ to keep the notation uncluttered.
This is a conjugate prior distribution for the Gaussian likelihood function given in Eq.~3.
By substituting the likelihood (Eq.~3) and prior distribution (Eq.~5) into Bayes' theorem (Eq.~4), 
it is easily checked that the posterior distribution also has a Gaussian-inverse-gamma form, and 
thus we obtain the relations for the hyperparameters (Eq.~6).

In the numerical simulations discussed in the main text, 
the hyperparameters in the prior distributions were initially set as follows.
We chose $\bm{\chi}^\mathrm{old}_i = \bm{0}$ and $\alpha^\mathrm{old}_i = \beta^\mathrm{old}_i = 0$, which correspond to the values for an uninformative prior distribution.
The covariance matrix $\Sigma_i^\mathrm{old}$ was initially chosen to be a diagonal matrix as
\[ \Sigma_i^\mathrm{old} = \mathrm{diag}[\lambda_i^{-1}, M_i\lambda_i^{-1},\dots,M_i\lambda_i^{-1}], \]
and we determined the precision parameters $\lambda_i$ by maximizing the marginal likelihood, just as we did for $M_i$ (see the next section).

\section{Approximated maximization of the marginal likelihood}
In the analyses presented in the main text, 
we determined the values of $M_i$ and $\lambda_i$ by maximizing the log marginal likelihood function $L_i$,
following a Bayesian model selection method.
However, because in general we cannot analytically optimize $L_i$ with respect to these 
parameters,
we approximated the optimal parameter values, $M_i^*$ and $\lambda_i^*$.
Specifically, we considered many points distributed over the plane $(M_i, \lambda_i)$ as
\[
\begin{cases}
	M_i &= 0,1,\dots,M_\mathrm{max},\\
	\log \lambda_i &= 0,1,\dots,10,
\end{cases}
\]
where $M_\mathrm{max}=10$ for the first example and $5$ for the second example.
For each point $(M_i, \lambda_i)$, we calculated the posterior distribution and the corresponding value of $L_i(M_i, \lambda_i)$.
We then found the point $(M_i^*, \lambda_i^*)$ that yielded the largest $L_i$ and used it as the optimal value.
We note that in Fig.~2d in the main text, we plotted the logarithm of $L_i[M_i, \lambda_i^*(M_i)]$,
where $\lambda_i^*(M_i)$ is the optimum under fixed $M_i$, i.e.,
$\lambda_i^*(M_i) \equiv \mathrm{argmax}_{\lambda_i} L_i(M_i,\lambda_i)$.

\section{Model equations used in the first example}
In the simulations whose results are plotted in Fig.~2 in the main text, 
we used the van der Pol-type oscillators given by
\begin{align*}
	\dot{x}_1 &= y_1 + K(x_2-x_1) + \xi_{x,1}(t),\\
	\dot{y}_1 &= \epsilon_1(1-x_1^2)y_1 - x_1 + Kx_2^2 y_2 + \xi_{y,1}(t),\\
	\dot{x}_2 &= y_2 - K x_1^2 y_1 + \xi_{x,2}(t),\\
	\dot{y}_2 &= \epsilon_2(1-x_2^2)y_2 - x_2 + K x_1 y_1^2 + \xi_{y,2}(t),
\end{align*}
with $\langle \xi_{a,i}(s)\xi_{b,j}(t) \rangle = \sigma^2\delta_{ij}\delta_{ab}\delta(s-t).$
Parameter values are $\epsilon_1 = 0.3, \epsilon_2=0.7, K=0.01$ and $\sigma = 0.03$.

\section{Neuron and synapse models used in the second example}
In the simulations whose results are plotted in Fig.~3 in the main text, for each excitatory neuron,
we used the Hodgkin-Huxley model \cite{Hodgkin:1952td}, given by
\begin{align*}
	C\dot{V} &= G_\mathrm{Na} m^3 h (E_\mathrm{Na}-V) + G_\mathrm{K} n^4 (E_\mathrm{K}-V) + G_\mathrm{L}(E_\mathrm{L}-V) + I_\mathrm{input} + \xi_V,\\
	\dot{m} &= \alpha_m(V) (1-m) - \beta_m(V) m + \xi_m, \alpha_m(V)=\frac{0.1(V+40)}{1-\exp[(-V-40)/10]}, \beta_m(V) = 4\exp\frac{-V-65}{18},\\
	\dot{h} &= \alpha_h(V) (1-h) - \beta_h(V) h + \xi_h,
	\alpha_h(V) = 0.07 \exp \frac{-V-65}{20}, \beta_h(V) = \frac{1}{1+\exp[(-V-35)/10]},\\
	\dot{n} &= \alpha_n(V) (1-n) - \beta_n(V) n + \xi_n,
	\alpha_n(V) = \frac{0.01(V+55)}{1-\exp[(-V-55)/10]}, \beta_n(V) = 0.125\exp\frac{-V-65}{80},
\end{align*}
with parameter values $C=1, G_\mathrm{Na}=120, G_\mathrm{K}=36, G_\mathrm{L}=0.3, E_\mathrm{Na}=50, E_\mathrm{K}=-77, E_\mathrm{L}=-54.4$.  
For each inhibitory neuron, we used a model of fast-spiking neurons \cite{Erisir:1999tn}, given by
\begin{align*}
	C\dot{V} &= G_\mathrm{Na} m^3 h (E_\mathrm{Na}-V) + G_\mathrm{K} n^2 (E_\mathrm{K}-V) + G_\mathrm{L}(E_\mathrm{L}-V) + I_\mathrm{input} + \xi_V,\\
	\dot{m} &= \alpha_m(V) (1-m) - \beta_m(V) m + \xi_m, 
	\alpha_m(V)=\frac{40(V-75)}{1-\exp[(75-V)/13.5]}, \beta_m(V) = 1.2262\exp\frac{-V}{42.248},\\
	\dot{h} &= \alpha_h(V) (1-h) - \beta_h(V) h + \xi_h,
	\alpha_h(V) = 0.0035 \exp \frac{-V}{24.186}, \beta_h(V) = \frac{0.017(-51.25-V)}{\exp[(-51.25-V)/5.2]-1},\\
	\dot{n} &= \alpha_n(V) (1-n) - \beta_n(V) n + \xi_n,
	\alpha_n(V) = \frac{V-95}{1-\exp[(95-V)/11.8]}, \beta_n(V) = 0.025\exp\frac{-V}{22.222},
\end{align*}
with parameter values $C=1, G_\mathrm{Na}=112, G_\mathrm{K}=224, G_\mathrm{L}=0.1, E_\mathrm{Na}=55, E_\mathrm{K}=-97, E_\mathrm{L}=-70.0$.

For each cell $i$, the input current was the sum of the bias and synaptic currents:
$I_{\mathrm{input},i} = I_{\mathrm{bias},i} + \sum_{j \in \mathrm{pre}_i} I_{\mathrm{syn},ij}$.
Here, pre$_i$ denotes the set of indices of the cells that send synaptic inputs to the $i$-th cell.
We set the bias currents as $I_{\mathrm{bias},i} = 30,32,6,6.5,34,36,38$ for $i=1,\dots,7$, respectively.
For each synaptic current, $I_{\mathrm{syn},ij}$, we adopted the kinetic synapse model \cite{Destexhe:1994uc} as
\[I_{\mathrm{syn},ij} = G_{ij} r_{ij}(t) [V_i(t)-E_{ij}]. \]
Here, $r_{ij}$ represents the fraction of bound receptor proteins.
Its dynamics are given by
\[ \frac{dr_{ij}}{dt} = \alpha_{ij} T_{ij} (1-r_{ij}) - \beta_{ij} r_{ij}, \]
where $T_{ij}$ denotes the concentration of the neurotransmitter, 
which is set to 1 when the presynaptic cell emits a spike and then reset to 0 after 1 millisecond.
The constants $\alpha_{ij}$ and $\beta_{ij}$ determine the timescale of the kinetics of $r_{ij}$,
$E_{ij}$ is the reversal potential (in millivolts),
and $G_{ij}$ is the synaptic conductance.
We used the values $(\alpha_{ij}, \beta_{ij}, E_{ij}, G_{ij}) = (1.1, 0.67, 0, 0.5)$ for excitatory and $(9.8, 0.2, -75, 0.4)$ for inhibitory synapses.
For each cell, a weak, independent noise function $\xi_{\cdot,i}$ was added to the membrane voltage $V_i$ and channel variables, $m_i,h_i$, and $n_i$.
The noise was a Gaussian white noise satisfying $\langle \xi_{x,i}(t) \rangle = 0$
and $\langle \xi_{x,i}(t)\xi_{y,j}(s) \rangle = \sigma_x^2 \delta_{xy} \delta_{ij} \delta(t-s)$,
where $x,y = V,m,h,n$, and $i$ and $j$ are the cell indices.
The noise strengths used are $\sigma_V = 0.5$ and $\sigma_m = \sigma_h = \sigma_n = 5 \times 10^{-6}$.

\section{Results with $N=20$ oscillatory neurons.}
\begin{figure}[h]
	\begin{center}
		\includegraphics[width=17cm]{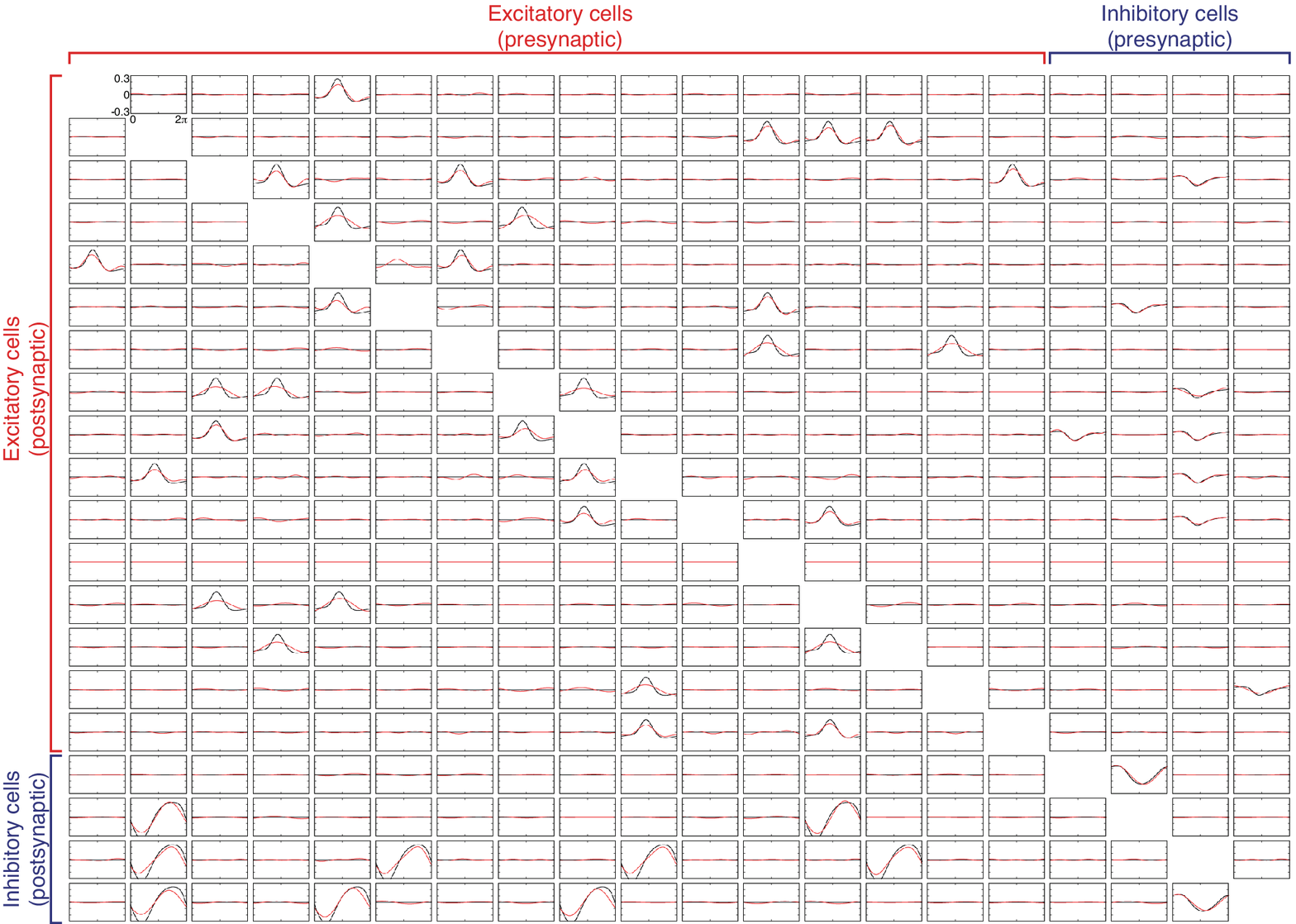}
	\end{center}
	\caption{
		Estimated (red trace) and true (black trace) coupling functions $\hat{\Gamma}_{ij}$ for the example with twenty neurons.
		For the estimated curves, the posterior mean is plotted.
		It should be noted that couplings between distinct types 
		(excitatory and inhibitory) of postsynaptic and presynaptic cells 
		lead to significantly different forms for the coupling functions.
	}
	\label{fig:SN-N20}
\end{figure}

To examine whether our method yields reasonable estimations for larger networks, 
we considered a network 20 spiking neurons.
The network consists of 16 Hodgkin-Huxley $(i=1,\dots,16)$ and 4 fast-spiking ($i=17,\dots,20)$ neurons.
The induced bias currents were $I_{\mathrm{bias},i} = 29 + i$ for the Hodgkin-Huxley and 
$I_{\mathrm{bias},i} = 6 + 0.25(i-17)$ for the fast-spiking neurons.
Unlike in the case of the second example considered in the main text, 
in the present case, all neurons were active in every trial.
All the parameter values were chosen to be the same as in the example with seven neurons, 
except that the parameter values were chosen as $\sigma_V = 0.1$ and $\sigma_m = \sigma_h = \sigma_n = 1 \times 10^{-6}$, and $G_{ij} = 0.1$ (0.08) for excitatory (inhibitory) cells.
Figure \ref{fig:SN-N20} plots the results of the estimation obtained using the voltage trace data,
which is five times the size of that in the example with seven neurons.
We find that the estimation is sufficiently good that the existence, directionality and heterogeneity 
of the couplings are all distinguishable, 
although fine details of the coupling functions are not completely captured.

\bibliography{paper}

\end{document}